\documentclass[prl,twocolumn,superscriptaddress,floatfix]{revtex4}                 
\usepackage{amsmath,amssymb}
\usepackage{amstext}
\usepackage{latexsym}
\usepackage[dvips]{graphicx}
\usepackage{color}

\newcommand{\beq}{\begin{equation}}
\newcommand{\eeq}{\end{equation}}

\begin{document}
\title{Quantum Billiards in Optical Lattices}

\author{Simone Montangero}
\affiliation{NEST-CNR-INFM \& Scuola Normale Superiore, Piazza dei
             Cavalieri 7, I-56126 Pisa, Italy}

\author{Diego Frustaglia}
\affiliation{Departamento de F\'isica Aplicada II, Universidad de Sevilla, 
             E-41012 Sevilla, Spain}

\author{Tommaso Calarco}
\affiliation{Institute for Quantum Information Processing, University of Ulm, 
             D-89069 Ulm, Germany}
\author{Rosario Fazio}
\affiliation{NEST-CNR-INFM \& Scuola Normale Superiore, Piazza dei
             Cavalieri 7, I-56126 Pisa, Italy}
\affiliation{International School for Advanced Studies (SISSA), via
             Beirut 2-4,  I-34014 Trieste, Italy}

\begin{abstract}
We study finite two dimensional spin lattices with definite geometry 
(spin  billiards) demonstrating the display of collective integrable or chaotic
dynamics depending on their shape. We show that such systems 
can be quantum simulated by ultra-cold atoms in optical lattices 
and discuss how to identify their dynamical features in a realistic experimental 
setup. Possible applications are the simulation of quantum information tasks in 
mesoscopic devices.
\end{abstract}

\maketitle

During the last decades, billiards have been the testbed of classical 
and quantum chaos~\cite{chaos,quantbill} due to their simplicity united to the richness of 
their displayed dynamics~\cite{billiards}.  Theoretical evidences of the quantum 
manifestation of chaos in billiards were first confirmed experimentally in the 
spectral statistic of microwave resonators~\cite{stockmann},   
quasi-two-dimensional superconducting resonators~\cite{graf}, and
atom-optic billiards \cite{davidson}. A very important step forward in the 
field occurred when it was realized that the properties of mesoscopic systems
could be very sensitive, under appropriate circumstances, to the integrability 
properties of the underlying classical model~\cite{richter-beenakker}. We recall, as an 
example, the study of conductance fluctuations in quantum dots~\cite{cmarcus}.
More generally, quantum billiards have shown to determine the dynamical properties 
of charge~\cite{chargetransp}, spin~\cite{spintransp} and entanglement~\cite{enttransp} 
in nanostructures. The physics associated to quantum billiards has been shown very 
recently to be relevant in the study of graphene~\cite{graphbill}.

The ongoing interest in the study of quantum chaos stimulates the search for new physical systems 
where it is possible to experimentally study complex dynamical behaviour. In this 
Letter we propose to realize quantum billiards using optical lattices, which 
have been proved to be an excellent arena to study quantum many-body 
systems~\cite{lewenstein}. 

Fundamental to our proposal is that optical lattices can operate as 
universal simulators~\cite{hamsim}, i.e., by means of an appropriate dynamical control
it is possible to reproduce the dynamics of any given spin Hamiltonian. Moreover, by means 
of the modelling of the form of the external trap it is possible to effectively define 
finite-size lattices.  The class of billiards defined in this work are finite two 
dimensional optical lattices of given geometry, 
where collective excitations propagate and interfere as they back-reflect against the 
geometrical boundaries of the lattice. We therefore talk about {\em spin billiards}. 
This class shows a rich set of possible configurations, serving as a model system for 
different implementations: Depending on the system size, boundary conditions, lattice
coordination number, and interaction Hamiltonian between the spins, one can either 
recover the known results on quantum billiards or model new physical systems showing 
original features. In the realm of cold atomic gases the distinction between 
regular and chaotic dynamics will appear in the momentum distribution of the atoms
or in the fluorescence signal. Present-day technology permits the simulation of 
the spin billiards introduced here. 

Optical lattices offer unique possibilities to simulate chaotic or integrable 
dynamics in a controlled way. The possibility to study billiards in this context gives a brand
new perspective to a classic, and well studied, problem; numerous
new questions can be addressed. On one side it is possible to explore the transition to 
chaos in a number of different spin-Hamiltonian depending of its symmetries. On the 
other side it is essential to understand the realization of the billiard, the measurement of relevant quantities and the sources of imperfections that may mask the physics we want to 
describe. We decide to first address this last topics. To this end we consider a model 
Hamiltonian which can be mapped on that of a particle hopping on a finite lattice 
(the billiard). We will use the language of spin 
as it is natural in this case and it applies also to those Hamiltonians where the mapping 
to a tight-binding model does not apply. 

\begin{figure}[t]
\begin{center}
A \includegraphics[scale=0.368]{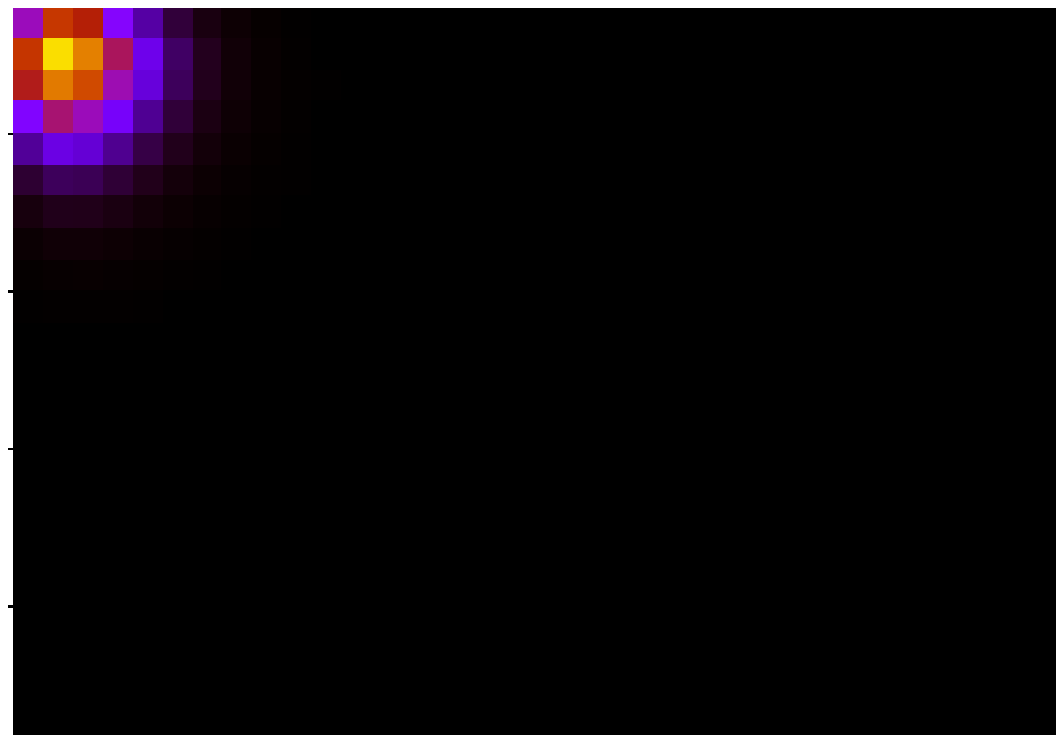}
B\includegraphics[scale=0.365]{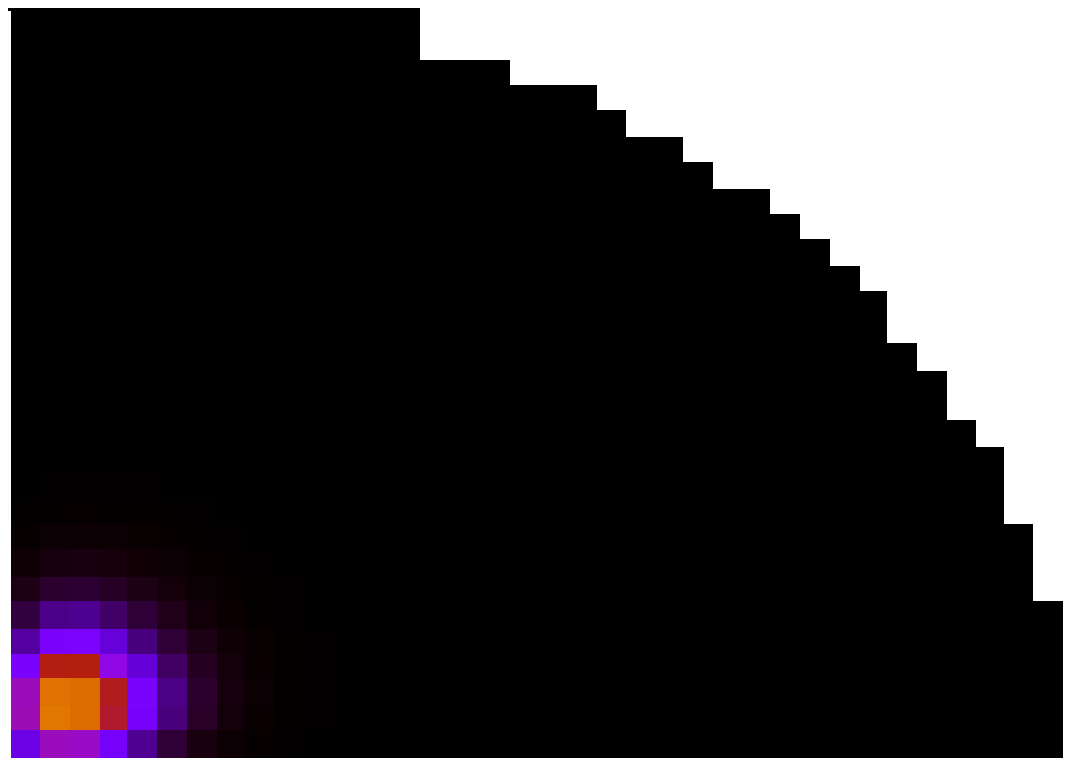}
C\includegraphics[scale=0.372]{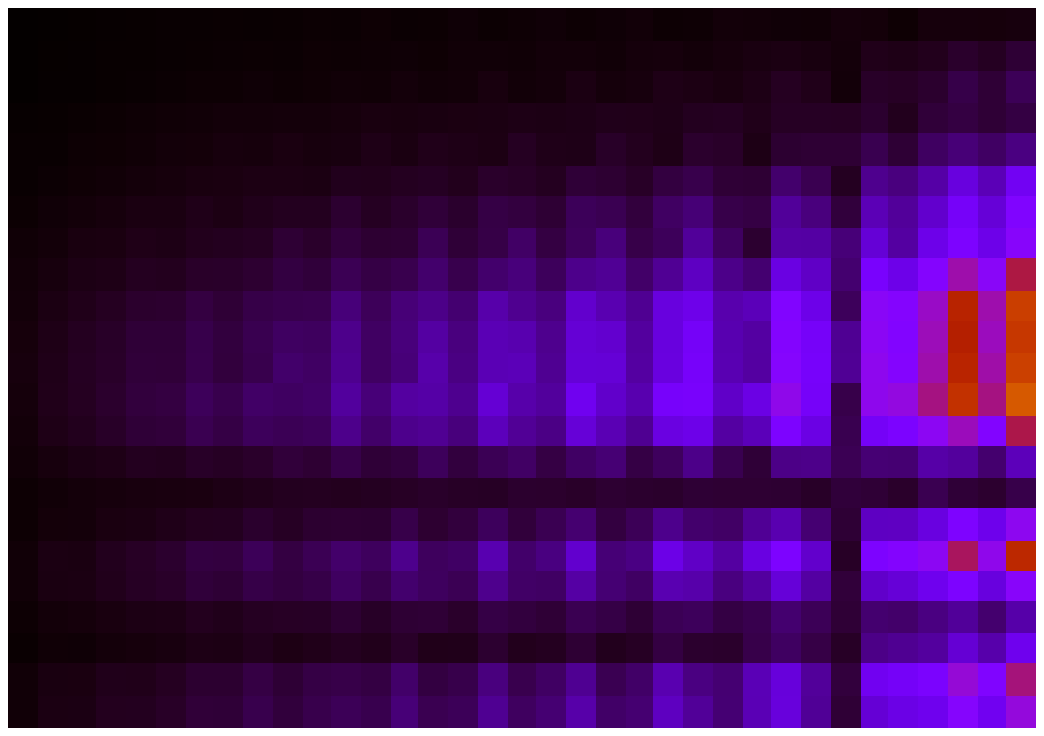}
D\includegraphics[scale=0.365]{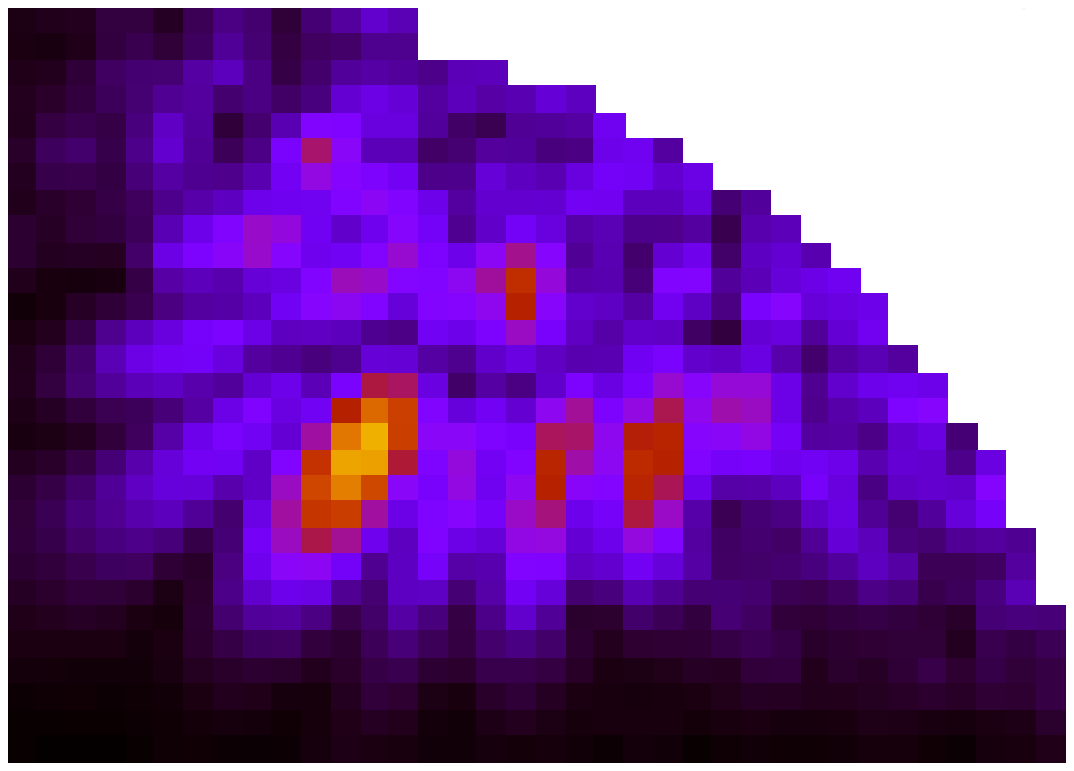}
E\includegraphics[scale=0.372]{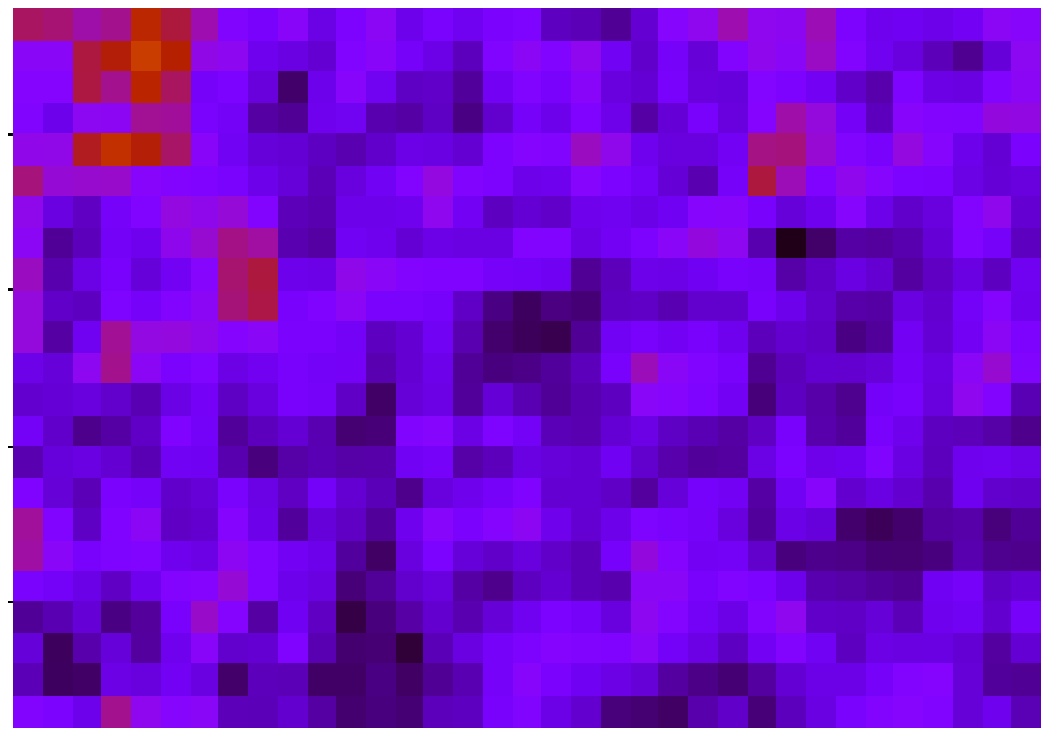}
F\includegraphics[scale=0.365]{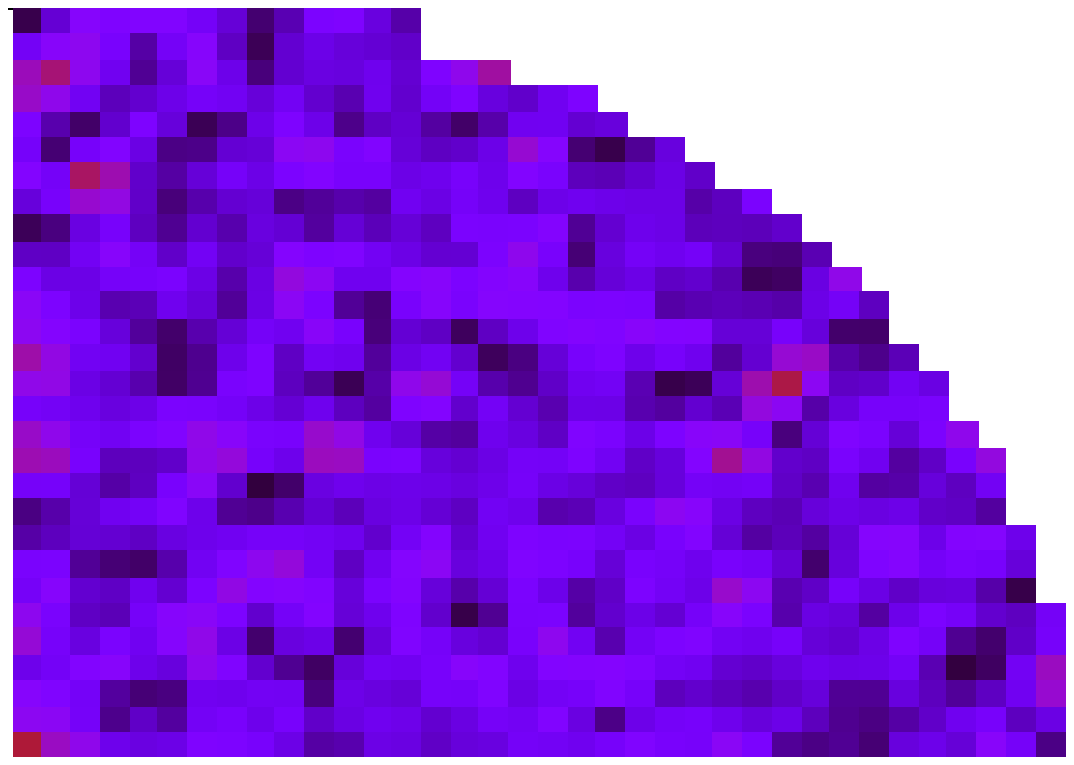}
\caption{Snapshots of the populations $|\psi_t|^2$ 
  time evolution in the rectangular billiard
  (left) and stadium billiard (right), for $t=0^+$ (uppermost figures),
  $t \gtrsim T_L/2$ (middle) and $t= 10 \, T_L$ (bottom). Color code goes
  from black (zero) to blue and red with increasing probability.} 
\label{timev}
  \end{center}
\end{figure}

{\em The model -} 
We consider a two dimensional $1/2$-spin lattice with
nearest-neighbor $XX$ interaction in a transverse magnetic field. 
The Hamiltonian reads
\beq
\mathcal{H} = \lambda \sum_{<m,m'>} (\sigma_x^m \sigma_x^{m'} + \sigma_y^m \sigma_y^{m'})
+ \sum_{m} \sigma_z^m
\label{ham}
\eeq
where $\sigma^i_\alpha$ are the Pauli matrices, $m= \mathcal{M}(i,j)$ 
is the composed index of the two dimensional qubits coding $\{i,j\}$,
and the sum $<m,m'>$ runs over nearest-neighbor spins on a square 
lattice with coordination number four (except at the boundaries) and 
free boundary conditions. We set $\lambda=1$ and
$\hbar=1$. As $\mathcal{H}$ commutes with the total magnetization, 
we restrict to the subspace with total magnetization equal to one (in this 
particular sector a mapping onto the single particle problem applies). 
The discrete space structure is a key-feature of optical lattices, we therefore
need to re-discuss the effect of different billiard
shapes on the dynamics. The first billiard under consideration is rectangular 
(billiard R, Fig.~\ref{timev}, left). We simulate the time evolution of a
wave function initially peaked in an angle (single spin flipped, see 
Fig.~\ref{timev}~A). This situation can be realized by starting from a
Mott insulator state with unit occupancy in a two-dimensional optical
lattice \cite{rmplatt}. Two atomic hyperfine levels serve as the two
pseudo-spin states, and all atoms can be prepared in one of the two by
optical pumping. A laser can then be used to excite atoms from
specific lattice sites to untrapped (continuum) states, removing them
from the lattice. Its focus can be swept all along the border of a
pre-defined region, leaving a regularly filled lattice of uniformly
polarized atoms corresponding to the chosen billiard shape. At this
point a Raman $\pi$-pulse on resonance with the transition between the
two hyperfine levels can be used to flip the state of one atom at one
of the billiard's vertices. In this particular geometry, shining the
vertex atom with the edge of the laser spot will suffice, eliminating
the need for subwavelength addressing. The simulation of the
propagation of the resulting spin wave can then take place following a
stroboscopic procedure based on lattice-driven state-dependent
collisions between neighboring atoms \cite{hamsim,molmer}. Billiard S
(Fig.~\ref{timev},  
right) is a quarter of a Bunimovich stadium and the initial condition 
is, as before, localized at a boundary angle (see Fig.~\ref{timev}~B). 
In both cases, the initial condition reads
\beq
\label{ic}
|\psi_{t=0} \rangle  \equiv |\mathcal{M}(0,0)\rangle. 
\eeq
Note that our analysis applies also to four-fold symmetric billiards, 
the symmetry axis of which passes through the site of the initial excitation 
$\mathcal{M}(0,0)$.
Our choice allows us to study the time evolution of an excitation
neglecting the effects of central and axial symmetries. 
The initial condition (\ref{ic}) is such that the dynamics will be 
influenced also by very high energy levels, i.e., the dynamics 
is very far from being composed only by the low-lying excitations. 
On the contrary, in the continuum limit 
and starting with a different initial condition as, e.g., a Gaussian packet, 
one would recover the usual physics of electronic billiards. Using the
lattice on one hand leads to a discretization of the space, altering
the geometric nature of curved edges (see Fig.~\ref{timev}B, D, F); on
the other hand, through the stroboscopic nature of the dynamic
simulation it allows to ``freeze'' the system at a very well defined
point in its time evolution for the purpose of state detection. 
The billiard dynamics is characterized by two time scales $T_L$ and $T_\lambda$:
The first one is related to the characteristic length of the billiards
$L$ ($\sim 30$ sites in our simulations), corresponding to the 
time needed for the first revival of excitations;
the second timescale is given by the time needed to perform a swap
between two neighboring spins, related to the inter-site coupling strength 
$T_\lambda = \pi/(4 \lambda)$. The relation of the two timescales is 
$T_L \propto 2L \, T_\lambda$.
Fig.~\ref{timev} C-F depict snapshots of the site excitation amplitude after 
time evolution at two different final times $t_f$ for billiards R and S:
For $t_f \gtrsim T_L/2$ the effect of different boundary shapes is already 
visible (Fig.~\ref{timev}~C and D). 
For longer times, $t_f \gg T_L$ the collective excitations spread all over
the billiards, showing irregular profiles with no distinguishable features 
at first sight (Fig.~\ref{timev}~E and F).  
However, for $t_f \propto n \, T_L$ ($n \in\mathbb{N}$), 
a large revival at the initial site is still visible for the rectangular billiard 
resulting from constructive interference, while this is no longer possible 
for the stadium billiard. 
These are the first signatures resembling chaotic and integrable dynamics 
in billiards realized in optical lattices. In the following we demonstrate that 
this is indeed the case and that its characteristic features can be detected and 
quantified  experimentally. 

\begin{figure}[t]
  \begin{center}
\includegraphics[scale=0.3]{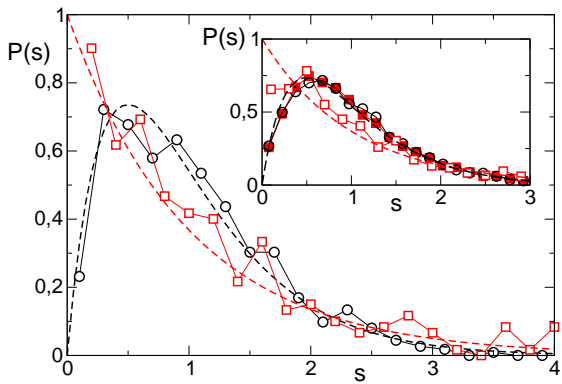}
\caption{Level spacing statistic for the billiard R (red squares) and billiard
  B (black circles). Dotted lines follow the Poisson (red) and Semi-Poisson
  statistics (black). Inset: LSS averaged over $N_R=10$ different configurations of
  defects with $P_D = 5 \cdot 10^{-3}$ (empty symbols)  $P_D = 5 \cdot  10^{-2}$ (full
  symbols). The black [red] dashed line
  follows the theoretical prediction $P(s)=4 s \exp(- 2 s)$ [$P(s)=\exp(-s)$].} 
\label{stat}
  \end{center}
\end{figure}
We first check the level spacing statistics (LSS) for the two billiards R and S.
Following Bohigas' conjecture we expect billiard R to show Poisson LSS, while 
billiard S should present something different due to the effect of level repulsion. 
Indeed, as shown in Fig.~\ref{stat}, we find that billiard R displays a well defined 
Poisson LSS (red squares). For billiard S, instead, we find a
Semi-Poisson LSS typical of semi-integrable systems (also appearing in the 
Anderson Metal-insulator transition)~\cite{bohigas-semiint}. The complete 
onset of chaos and the appearance of a Wigner-Dyson distribution is probably 
prevented due to the significant role still played by periodic orbits. 
A better convergence to the theoretical distribution can be obtained 
by considering defects, i.e., empty sites (see below). This
allows a better statistics by averaging over $N_R$ different 
configurations of defect probabilities ($P_D=5 \cdot 10^{-2}, 5 \cdot
10^{-3}$), as shown in the inset of Fig.~\ref{stat}.
Eventually, also billiard R displays Semi-Poisson statistics due to 
the presence of defects (inset of Fig.~\ref{stat}, red full squares).

\begin{figure}[t]
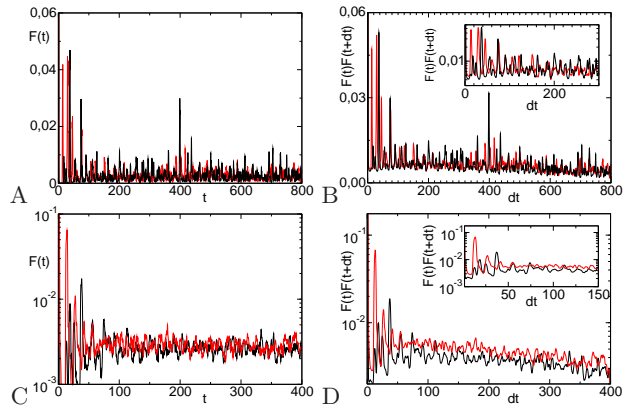

  \begin{center}
A\includegraphics[scale=0.15]{fig3a.eps} 
B\includegraphics[scale=0.15]{fig3b.eps}
C\includegraphics[scale=0.15]{fig5a.eps}
D\includegraphics[scale=0.15]{fig5b.eps}

\caption{Coarse grained Fidelity  $F_3$ (A,C) as a function of time for the
  integrable (black) and chaotic billiard (red) and the auto-correlation
  of the signals in the left figure (B,D) for $P_D=\epsilon=0$
  (upper panels) and $P_D= 5 \cdot 10^{-3}, \epsilon=10^{-5}$,
  $N_R=10$ (lower panels).  Insets: Magnification of the bigger figure.
}
\label{fid}
  \end{center}
\end{figure}

The striking differences in the LSS discussed above are reflected in 
other features of the spin-billiard dynamics that can be measured experimentally, as we show
hereafter. We consider in particular the momentum distribution and the fluorescence signal, as used for instance to detect single ions \cite{rmpions}. After 
its introduction by Peres~\cite{peres}, the survival probability or Fidelity $F$
\beq
F= | \langle \psi_{t=0} | \psi_t \rangle |^2,
\eeq
has been 
very useful to characterize the transition to chaos~\cite{qfid}. Here the main issue, due to the lattice spacing coinciding with optical wavelengths, is single-atom spatial resolution. Therefore, we consider a Coarse Grained Fidelity (CGF) defined as
\beq
F_n = | \langle \sum_{<i,j>}^{<i+n,j+n>}  \mathcal{M}(i,j) | \psi_t \rangle |^2,
\eeq
i.e., the survival probability in a
square region around the site $\{i,j\}$. The CGF can be obtained via 
fluorescence measurements in optical lattices without single site 
addressing, which is at the edge
of present day technology \cite{saddress}. Even if the CGF fails in 
detecting the finest details of the dynamics it still captures the main
differences between the integrable and chaotic systems. 
In Fig.~\ref{fid}~A and C we show the CGF ($n=3$) decay in the rectangular
(black) and stadium (red) billiards for different disorder settings together with 
their corresponding auto-correlation functions (Fig.~\ref{fid}~B and D). 
In spite of the random noise, both the CGF (Fig.~\ref{fid}~C) and its
auto-correlation (Fig.~\ref{fid}~D) reveal the fundamental time scale $T_L$.
Striking differences appear between R and S billiards: Periodic oscillations 
persist up to times of the order of $t_f \gg T_L$ in the integrable case (black), 
while in the chaotic case (red) a damping shows up on a time scale $t \gtrsim T_L$
revealing a rapid decay of correlations. 

\begin{figure}[t]
  \begin{center}
A\includegraphics[scale=0.3]{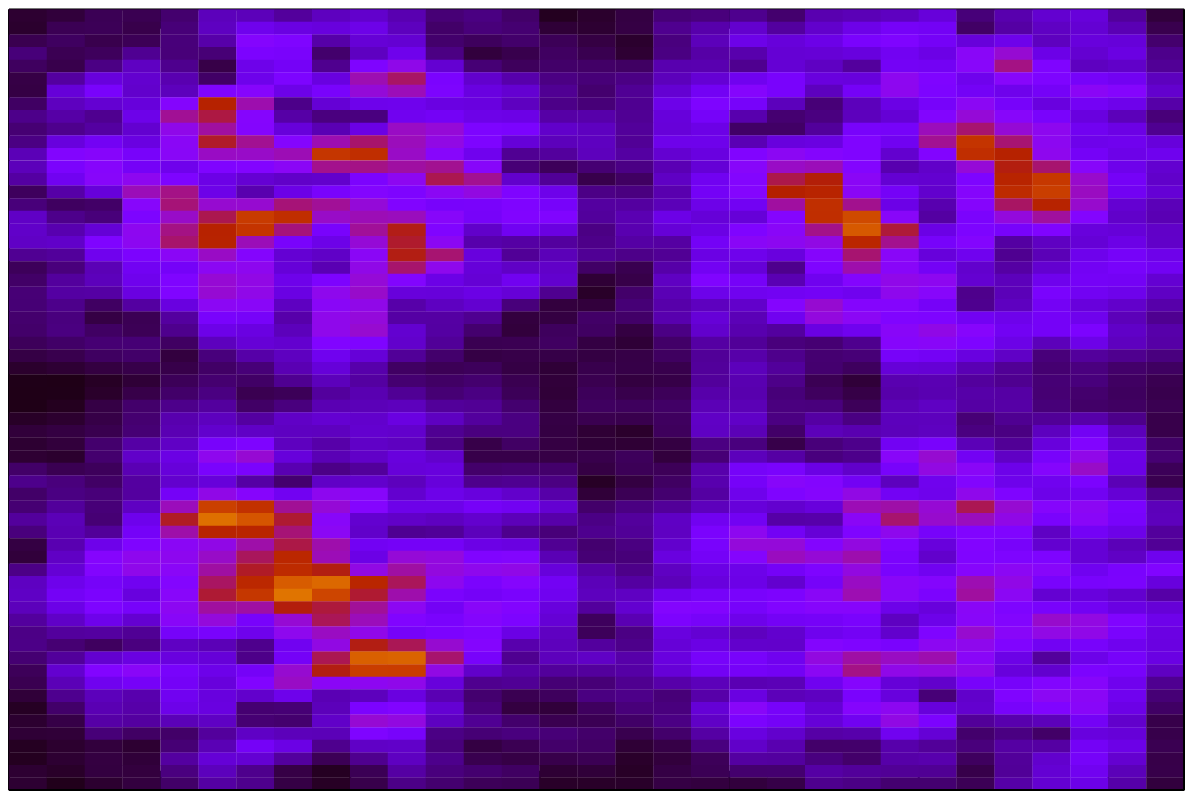}
B\includegraphics[scale=0.3]{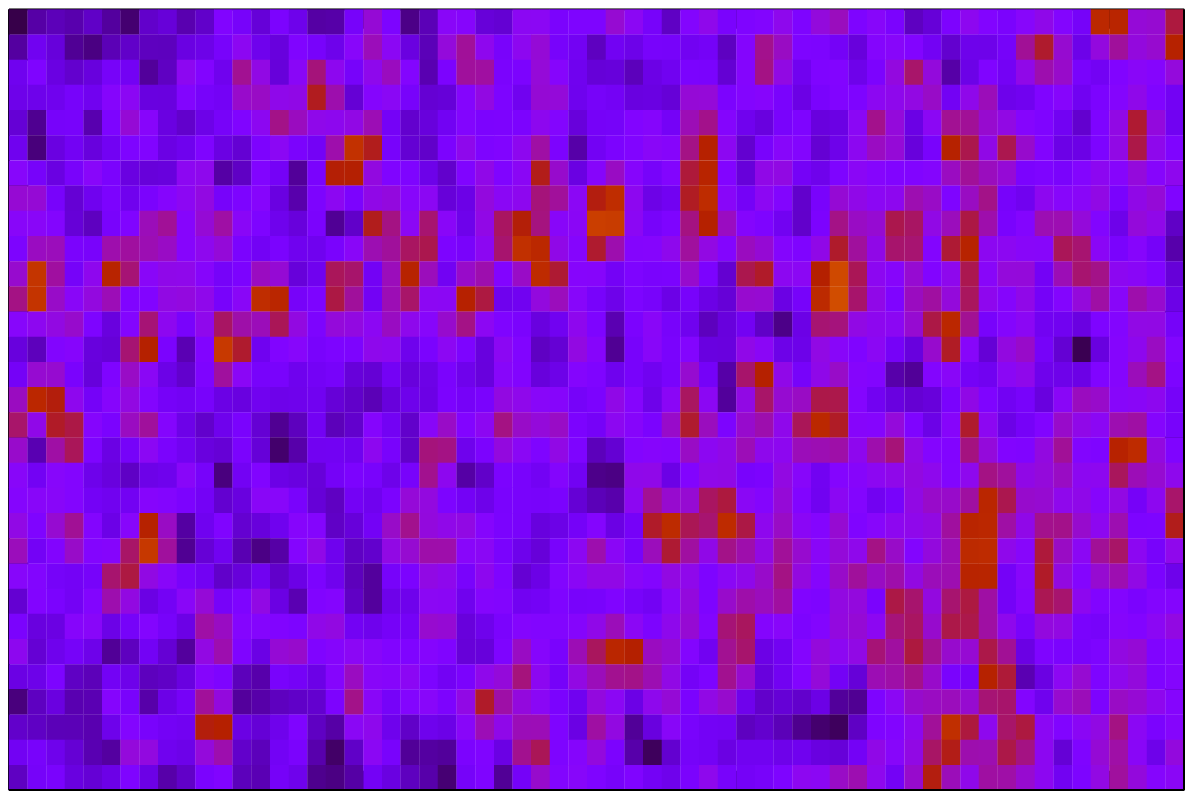}
C\includegraphics[scale=0.3]{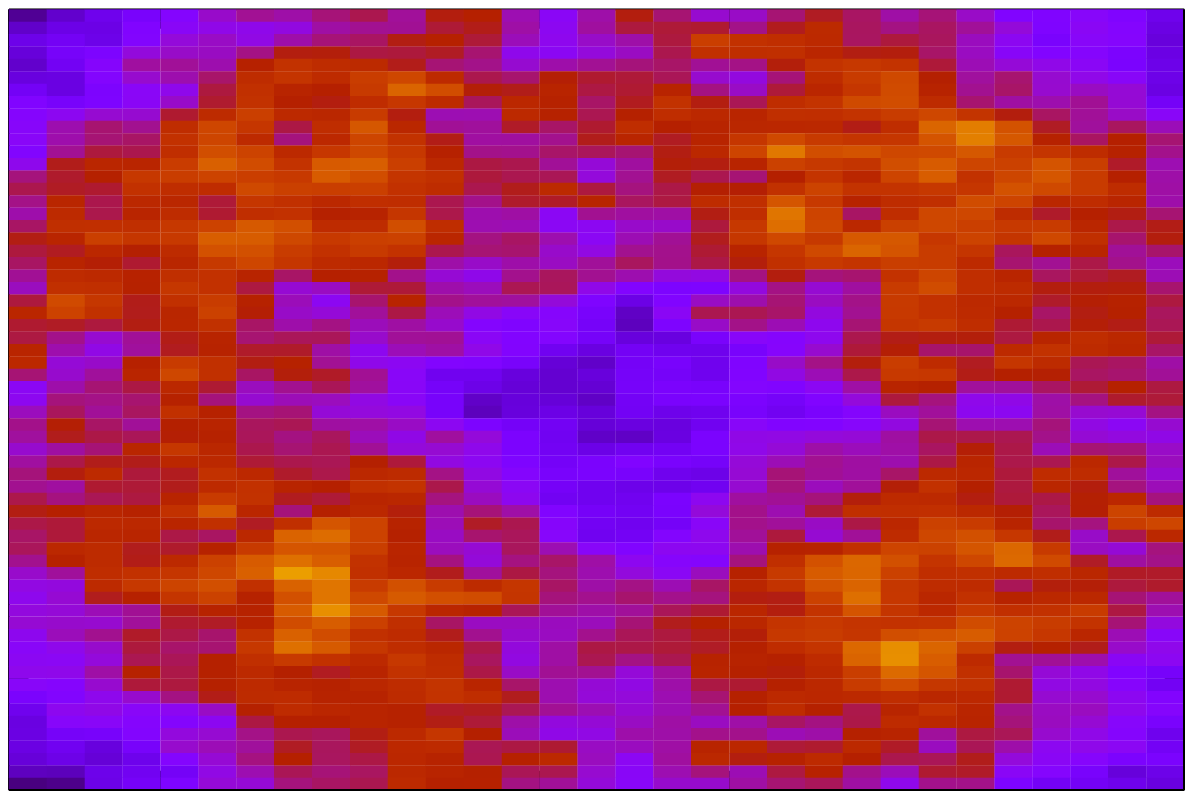}
D\includegraphics[scale=0.3]{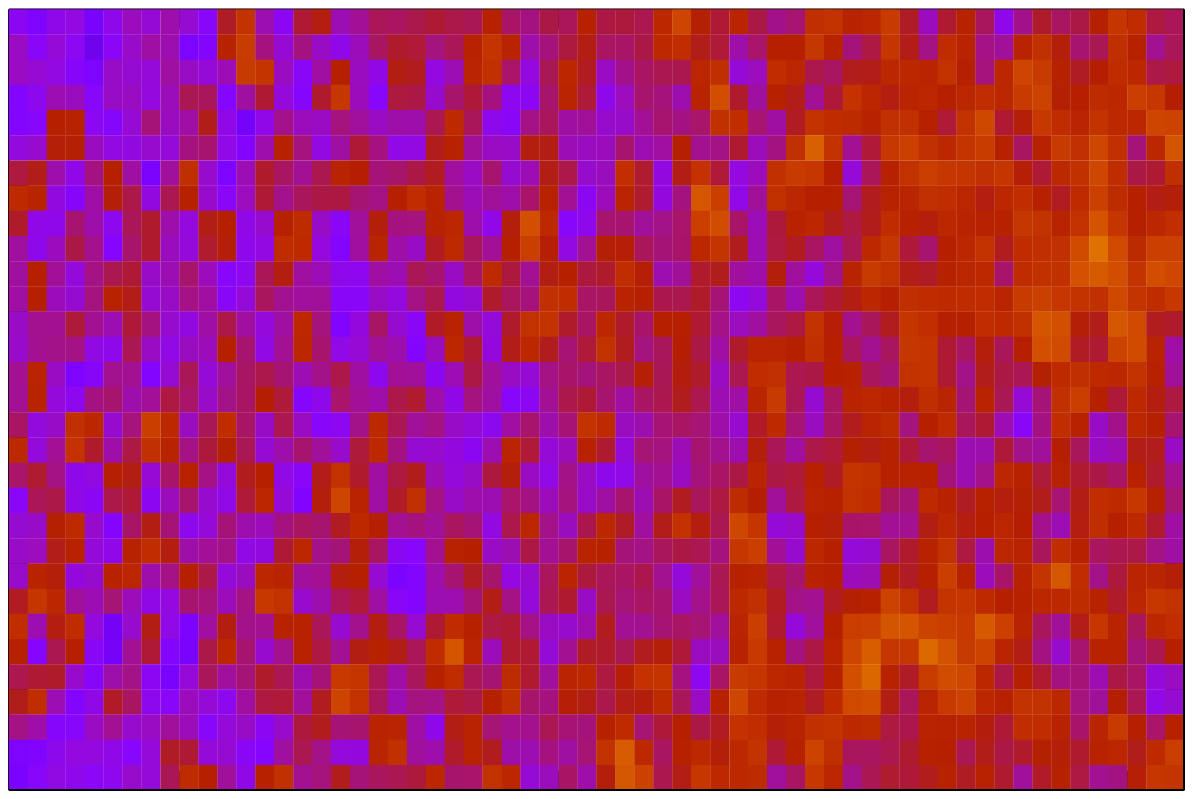}
E\includegraphics[scale=0.3]{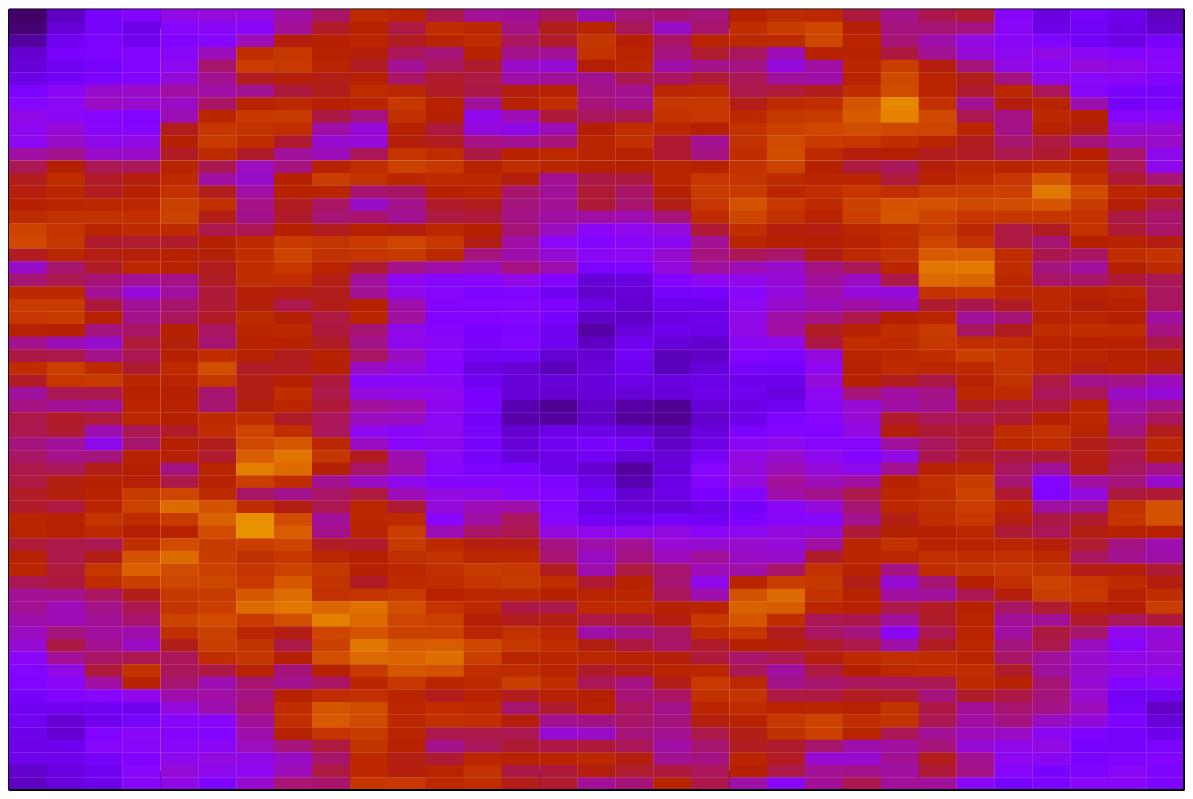}
F\includegraphics[scale=0.3]{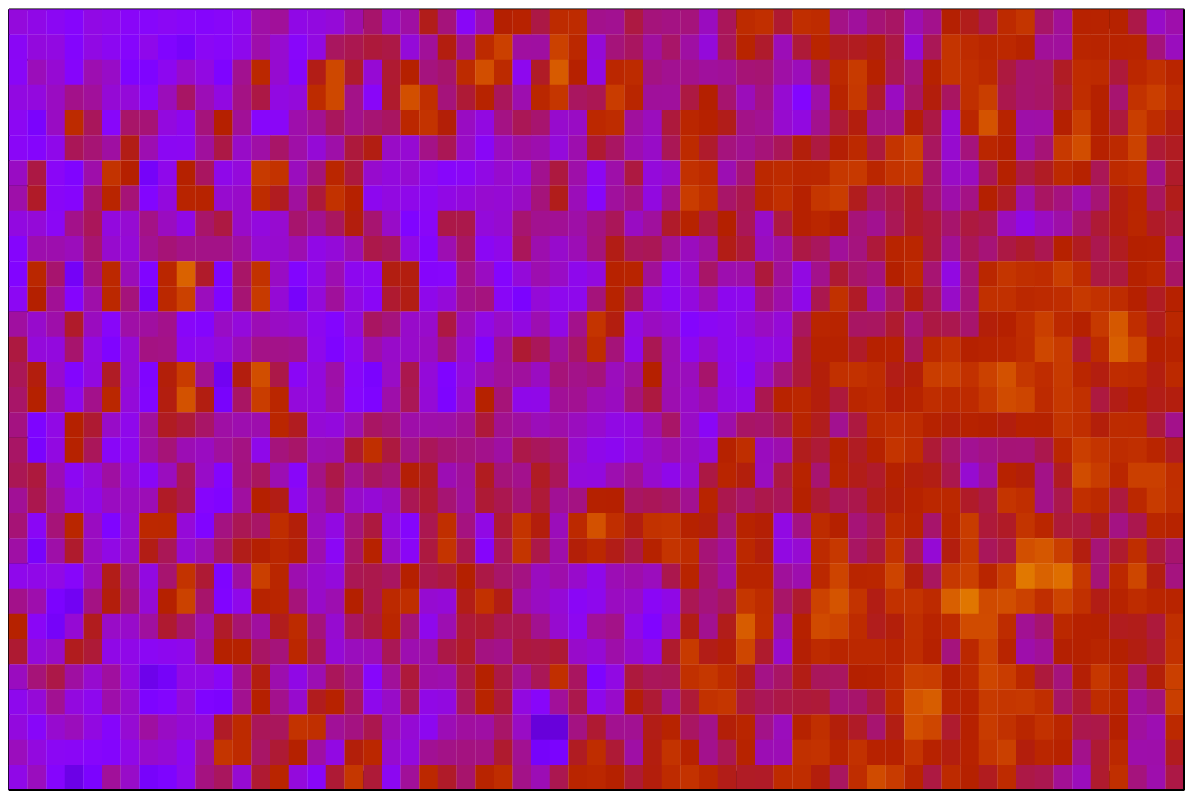}

\caption{2D FT of the Wave function $F(\omega_x,\omega_y)$ for the
  rectangular (left) and stadium (right) billiards at time $t_f= 10 \,
  T_L$ for $P_D=\epsilon=0$ (A,B); $P_D=5 \cdot 10^{-3}, 
  \epsilon=10^{-5}, N_R=10$ (C,D); and  $P_D=10^{-2},\epsilon=10^{-5},
  N_R=10$ (E,F). Color code is the same of Fig.~\ref{timev}.
} 
\label{fft2d}
  \end{center}
\end{figure}

Finally, we investigate the momentum distribution of the magnetization
in the two billiards at a final time $t_f$: The results are reported
in Fig.~\ref{fft2d}~A and B for the rectangular (left) and stadium 
billiards (right). Again the results show striking differences: In the
integrable case the number of frequencies relevant to the wave 
function are much less than in the chaotic case.
Moreover in the former case there is a structure (even though
quite complex) in the spectrum that is absent in the latter one.

{\em Experimental Implementation -}
As mentioned in the introduction spin billiards can be studied experimentally 
in optical lattices following the
idea of the Universal Quantum Simulator~\cite{hamsim}. 
The measurements of the CGF and momentum distribution can be performed via 
fluorescence and time-of-flight methods respectively~\cite{rmpions,rmplatt}. 
The signatures of chaos we are interested in arise from the evolution
of a single-spin excitation corresponding to single-particle signals. In order 
to attain a sufficient atom number resolution to detect it, an average over different 
realizations is required. Under ideal conditions, since the evolution is fully
deterministic, this should not be a problem and the result would be
fully reproducible. However, in a real experiment, errors might
introduce differences between repetitions which may result in a 
corrupted output. 
The major error sources are two: (i) imperfections in the realization of a 
square lattice with hundreds of sites with uniform occupation number one, 
and (ii) side effects of the parabolic magnetic field trapping the atoms in the 
studied region of the optical lattice~\cite{experr}. We can model these errors via 
the presence of defects or ``holes'' in the spin billiard (missing atom in a
optical lattice site) and errors in the gates performed to simulate
the dynamics. The Hamiltonian (\ref{ham}) is then replaced by
\beq
\mathcal{H}_1 =  \lambda \sum_{<m,m'>} (\sigma_x^m \sigma_x^{m'} + \sigma_y^m \sigma_y^{m'})
+ \sum_{m} \epsilon(t) (i+j) \sigma_z^m,
\label{gate1}
\eeq
where $m=\mathcal{M}_H(i,j)$ takes into accounts the presence of
defects and $\epsilon(t)$ fluctuates in $[0,\epsilon]$
with flat distribution. We repeat the previous analysis accounting for 
experimental errors with typical values  $P_D= 5 \cdot 10^{-3}$ and
$\epsilon= 10^{-5}$ averaging over $N_R$ different configurations 
\cite{experr}.
Fig.~\ref{fid} (lower panels) shows the coarse grained 
fidelity $F_3$ and the auto-correlation as a function of time R and S 
billiards in presence of experimental errors. In this case,
it is more difficult to distinguish between integrable and chaotic dynamics; 
however, a careful inspection can still reveal differences. As before, in the
integrable case the auto-correlation revivals last for longer times $t_f \gg
T_L$ and their visibility is greater than in the chaotic case. 
More clear signatures are found again in the momentum distribution
(Fig.~\ref{fft2d}, lower panels): The structures in the frequency 
domain lasts for very long times in the integrable billiard 
(even if slightly blurred) while they disappear in the chaotic case. 

Finally we would like to highlight the possible developments along the
lines presented here: The study and simulation of weak localization, 
quantum hall effect, disorder effects, quantum information protocols,
entanglement dynamics, and the role of different Hamiltonian and/or 
parameter regimes.
We also point out that a similar analysis could be performed
for alternative experimental setups as, e.g., lattices of coupled 
cavities \cite{martin}.
 
We thank M. Greiner and I. Bloch for insightful discussions.
This work was supported by the EC grants SCALA and EUROSQIP,  
by the ``Ram\'on y Cajal" Program of the Spanish Ministry of Education and Science, 
and by the Excellence Project P07-FQM-3037 of the Andalusian Government.


\begin{thebibliography}{99}

\bibitem{chaos} 
  see e.g.: Proceedings of the Les Houches Summer School on 
  {\em Chaos and Quantum Physics}, Les Houches, 1989, M. Giannoni, 
  A. Voros, and J. Zinn-Justin Eds. (North-Holland, Amsterdam, 1991);
  Proceedings of the International School of Physics "Enrico Fermi" on 
  {\em Quantum chaos}, Course CXIX, Varenna, 1991, G. Casati, I. Guarneri, 
  and U. Smilansky Eds. (North-Holland, Amsterdam, 1993).
\bibitem{quantbill} 
  M.V. Berry and M. Tabor  Proc. Roy. Soc. London A {\bf 356} 375 (1977);
  S.W. McDonald and A. N. Kaufman,  Phys. Rev. Lett. {\bf 42}, 1189 (1979); 
  G. Casati, B.V. Chirikov, and I. Guarnieri, {\it ibid.}  {\bf 54}, 1350 (1985); 
  R.A. Jalabert, H.U. Baranger and A.D. Stone, {\it ibid.} {\bf 65}, 2442 (1990);  
  G. Casati and T. Prosen, Physica D {\bf 131} 293 (1999).
\bibitem{billiards} 
  see e.g.: L. Reichl, {\em The Transition to Chaos}, Springer-Verlag (2004).
\bibitem{stockmann} 
  J. Stein and H-J. St\"ockmann, Phys. Rev. Lett. {\bf 64}, 2215 (1992).
\bibitem{graf} 
  H-D. Graf {\it et al.}, Phys. Rev. Lett. {\bf 69}, 1296 (1992).
\bibitem{davidson} 
  N. Davidson {\it et al.}, Phys. Rev. Lett. {\bf 74}, 1311 (1995). 
\bibitem{richter-beenakker}
  K. Richter, {\it Semiclassical Theory of Mesoscopic Quantum Systems} 
  (Springer-Verlag, Berlin, 2000); 
  C.J.Beenakker Rev. Mod. Phys. {\bf 69} 731 (1997).
\bibitem{cmarcus}  
  C.M. Marcus {\it et al.}, Phys. Rev. Lett. {\bf 69}, 506 (1992). 
\bibitem{chargetransp} 
  C.M. Marcus {\it et al.}, Chaos {\bf 3}, 643 (1993); 
  H.U. Baranger, R.A. Jalabert, and A.D. Stone, Chaos {\bf 3}, 665 (1993).
\bibitem{spintransp} 
  D. M. Zumb\"uhl {\it et al.}, Phys. Rev. Lett. {\bf 89}, 276803 (2002); I. L. Aleiner and 
  V. I. Fal'ko, {\it ibid.} {\bf 87}, 256801 (2001); O. Zaitsev, D. Frustaglia, 
  and K. Richter, {\it ibid.} {\bf 94}, 026809 (2005); {\it ibid.}, 
  Phys. Rev. B {\bf 72}, 155325 (2005).
\bibitem{enttransp} 
  C. W. J. Beenakker {\it et al.},
  in "Fundamental Problems of Mesoscopic Physics", edited by I. V. Lerner, B. L. 
  Altshuler, and Y. Gefen, NATO Science Series II vol. 154 (Kluwer, Dordrecht, 2004); 
  D. Frustaglia, S. Montangero, R. Fazio, Phys. Rev. B {\bf 74}, 165326 (2006); 
  J.H. Bardarson and C.W.J. Beenakker, {\it ibid.} {\bf 74}, 235307 (2006); 
  V.A. Gopar and D. Frustaglia, {\it ibid.} {\bf 77}, 153403 (2008).
\bibitem{graphbill} 
  F. Miao {\it et al.}, Science {\bf 317}, 1530 (2007).
\bibitem{lewenstein}
  M. Lewenstein {\it et al.}, Adv. Phys. {\bf 56}, 243 (2007).
\bibitem{hamsim} 
  E. Jan\'e {\it et al.}, Quantum Inf. and Comp. {\bf 3}, 15 (2003);  
  J.J. Garcia-Ripoll, M.A. Martin-Delgado, and J.I.Cirac, 
  Phys. Rev. Lett. {\bf 93}, 250405 (2004).
 \bibitem{rmplatt}
I. Bloch, J. Dalibard, and W. Zwerger, Rev. Mod. Phys. (in press), arXiv:0704.3011.
 \bibitem{molmer}
 A. S\o rensen and K. M\o lmer, Phys. Rev. Lett. {\bf 83}, 2274 (1999). 
\bibitem{bohigas-semiint} 
  O. Bohigas, in "Chaos and Quantum Physics",
  Proceedings of the Les Houches Summer School (1989), 
  M. J. Giannoni, A. Voros, and J. Zinn-Justin Eds.
  (Elsevier, New York, 1991);
  E. B.  Bogomolny,  U. Gerland, and C. Schmit,   
  Phys. Rev. E. {\bf  59}, R1315, (1999).
 \bibitem{rmpions}
 D. Leibfried {\it et al.}, Rev. Mod. Phys. {\bf 75}, 281 (2003).
\bibitem{peres} 
  A. Peres, Phys. Rev. A {\bf 30}, 1610 (1984).
\bibitem{qfid}  
  H.M. Pastawski, P.R. Levstein, and G. Usaj, 
  Phys. Rev. Lett. {\bf 75}, 4310 (1995);
  P. Jacquod, P.G. Silvestrov and C.W.J. Beenakker,
  Phys. Rev. E {\bf 64} 055203 (2001); 
  R. A. Jalabert and H. M. Pastawski, 
  Phys. Rev. Lett. {\bf 86}, 2490 (2001);
  G. Benenti {\it et al.}, 
  {\it ibid.} {\bf 87}, 227901 (2001); 
  F.M. Cucchietti {\it et al.},  
  {\it ibid.} {\bf 91} 210403 (2003). 
\bibitem{saddress}
A.V. Gorshkov {\it et al.},
Phys. Rev. Lett. {\bf 100}, 093005 (2008).
\bibitem{experr}
M. Greiner, private communication.
\bibitem{martin} 
M.J. Hartmann, F.G.S.L. Brand\~ao,  and M. B. Plenio, Nat. Phys.  {\bf 2}, 849 (2006).
\end{thebibliography}
\end{document}